\documentclass[aps,prl,amsmath,amssymb,reprint,showkeys,showpacs]{revtex4-1}

\newcommand{\del}{\partial}
\newcommand{\nn}{\nonumber}

\newcommand{\dd}{\mathrm{d}}
\newcommand{\be}{\begin{eqnarray}}
\newcommand{\ee}{\end{eqnarray}}

\newcommand{\bs}{\begin{subequations}\begin{eqnarray}}
\newcommand{\es}{\end{eqnarray}\end{subequations}}
\newcommand{\ba}{\begin{array}}
\newcommand{\ea}{\end{array}}
\newcommand{\bi}{\begin{itemize}}
\newcommand{\ei}{\end{itemize}}
\newcommand{\bei}{\begin{enumerate}}
\newcommand{\eei}{\end{enumerate}}
\newcommand{\lan}{\langle}
\newcommand{\ran}{\rangle}

\usepackage{graphicx}
\usepackage{bm}

\usepackage{subfigure}
\begin{document} 


\title{Fractional Einstein relation for strongly disordered
semiconductors}

\author{Takeshi Egami}\email{egami.takeshi@canon.co.jp}
\author{Koshiro Suzuki}\email{suzuki.koshiro@canon.co.jp} 
\author{Katsuhiro Watanabe}\email{watanabe.katsuhiro@canon.co.jp} 
\affiliation{Analysis Technology Development Center, Canon Inc., 30-2
Shimomaruko 3-chome, Ohta-ku, Tokyo 146-8501, Japan}

\date{\today}
%
\begin{abstract}
A novel Einstein relation (fractional Einstein relation, FER) for the
electric conduction in non-crystalline semiconductors is presented.
%
%
FER and the generalized Einstein relation (GER) [Phys. Rev. E {\bf8},
1296 (1998)] are compared to the result of the Monte Carlo (MC)
simulation, and is confirmed that FER exhibits better agreement than
GER.
The cruial feature of FER is that it reflects the violation of the
 detailed balance in the coarse-grained hopping process, while it is
 preserved in the original Einstein relation or GER.
%
%
%
\end{abstract}
\pacs{05.40.Fb,05.60.-k,02.50.Ey}
\maketitle 
{\it Introduction.--}\
Fluctuation-dissipation theorem (FDT) is one of the most fundamental
principles in statistical mechanics.
A significant example of FDT in kinetics is the Einstein relation (ER)
\cite{AnnPhys.17.549}, discovered in the Brownian motion.
The generic form of ER is given by $D = \mu k_{B} T$, where $T$ is the
temperature and $k_{B}$ is the Boltzmann constant.
Here, $D$ is the diffusion coefficient (fluctuation), and $\mu = v/F$ is
the mobility (dissipation), where $v$ is the steady drift velocity and
$F$ is the external force.
Explicit examples of ER can be found in various situations.
For instance, the Stokes-Einstein relation, $D = k_{B}T/(6\pi \eta d)$,
where $d$ the diameter of a sphere immersed in a fluid and $\eta$ the
viscosity, is well known.

For charged particles (carriers) in semiconductors, ER reads
\begin{align}
\frac{D}{\mu}
=
\frac{k_BT}{q},
\label{13101601}
\end{align}
where $q$ is the electric charge of the particle.
Such a relation is crucial for understanding the collective behavior of
carriers, because it is hard in general to measure the diffusion
coefficient, while it is relatively easy to measure the mobility.
However, it is known that ER does not hold in non-crystalline materials,
neither under equilibrium
\cite{ApplPhysLett.80.1948,ApplPhysLett.83.1998,PhysRevLett.94.036601,ApplPhysA.83.305}
nor nonequilibrium
\cite{PhysRevLett.63.547,JChemPhys.94.8276,Bassler,PhysRep.195.127,JNon-CrystSolids.198.214,JNon-CrystSolids.227.158}
conditions.
Roichman et al. \cite{ApplPhysLett.80.1948,ApplPhysLett.83.1998} have
proposed a modification of ER for the equilibrium case in terms of the
density of states (DOS).  
Instead of Eq.~(\ref{13101601}), they have postulated the following
relation,
\begin{align}
\frac{D}{\mu}
=
\frac{p}{q\frac{\del p}{\del \eta}},
\label{13101602}
\end{align}
where $p$ is the particle concentration and $\eta$ is the chemical
potential.
The particle concentration $p$ is expressed in terms of DOS, which we
denote $g(\varepsilon)$ with $\varepsilon$ the energy, and the
Fermi-Dirac distribution, $f(\varepsilon,\eta)$, as
$p=\int_{-\infty}^\infty d\varepsilon
g(\varepsilon)f(\varepsilon,\eta)$.

Establishing a valid ER for nonequilibrium cases has been partially
accomplished by Barkai et
al. \cite{PhysRevE.58.1296,PhysRevE.63.046118,PhysRep.195.127}. 
They have focused on the facts that the hopping conductance
\cite{MA,AHL,Bassler} is the dominant mechanism of the electric
conduction in disordered materials, and collective behaviors of carriers
show anomalous diffusion-advection \cite{Xerox1,Xerox2}, which is
believed to be described by the continuous time random walk (CTRW)
\cite{PhysRep.339.1}.
They have shown that CTRW is further described by the fractional
Fokker-Planck equation \cite{PhysRevE.61.132} under weak external
fields, and the generalized Einstein relation (GER),
\begin{align}
\langle x^2(t)\rangle_0=2\frac{k_BT}{F}\langle x(t)\rangle_F,
\label{GER}
\end{align}
holds under the assumption that the anomalous exponent of the waiting
time of the system with and without a driving force, $\alpha_F$ and
$\alpha_0$, are the same.
Here, $\langle x^2(t)\rangle_0$ is the mean-square displacement of the
carriers in the absence of an external field, $\langle x(t)\rangle_F$ is
the mean displacement in the presence of an external field, and $F$ is
the external force exerted on the carriers.
A representative case where GER is valid can be found in actin networks
\cite{PhysRevLett.77.4470,PhysRevLett.81.1134}.
However, although Barkai \cite{PhysRevE.63.046118} has conjectured that
GER would hold for the hopping conduction, quantitative comparisons of
GER with the experimental results or simulations have not been
performed.

In this letter, we propose a novel ER valid for the electric conduction
of non-crystalline semiconductors, which we refer to as the ``fractional
Einstein relation (FER)''.
For illustration, we consider the ``disorder model'' \cite{Bassler} of
the hopping conduction, where it is assumed that the electric conduction
is dominated by the static energy disorder of the hopping sites.
We compare FER and GER with the results of the Monte Carlo (MC)
simulation, and confirm that FER exhibits good agreement with MC, while
GER does not.

Note that the ``disorder model'' we consider is a well established
model, which is one of the two major microscopic models of the hopping
conductance. 
The other is the ``polaron model'' \cite{FKBN01,AnnPhys.8.343,SMT},
where it is assumed that the electric conduction is dominated by the
strong electron-phonon coupling.
To show the validity of FER in the ``polaron model'' is a future task,
but we believe that FER also holds in this model.
Note also that the representative phenomenological models of the hopping
conductance, namely the "multiple trapping model (MTM)"
\cite{SolidStateCommun.37.49,TMMA,MTM}, and the "Scher-Montroll model
(SMM)" \cite{Xerox1,Xerox2,SL1,SL2}, which describe the experimental
results of the time-of-flight (TOF) signals \cite{Xerox1,Xerox2}, are
the coarse-grained variants of the aforementioned microscopic models.

The letter is organized as follows.  
First we explain the disorder model and derive FER analytically.
Next, we compare FER and GER with MC simulation, and demonstrate that
FER is in good agreement with MC simulation, while GER is not.
Then we discuss the relation of our results to the previous studies.
Finally, we summarize our results.

\vspace{1em}
{\it Theory.--}\ 
We start with the introduction of the ``disorder model'' \cite{Bassler}
for the hopping conductance, which is essentially equivalent to the one
considered in Refs.~\cite{ESW,ESW2}.
The two crucial ingredients of the model are (i) the probability
distribution of the energy difference of the hopping sites and (ii) the
hopping rate of the carriers.

The distribution of the energy difference, which we denote
$h(\epsilon_{ij})$ with $\epsilon_{ji}\equiv \epsilon_{j}-\epsilon_{i}$
the energy difference of site $i$ and $j$, is determined by the DOS,
$g(\epsilon)$.
Note that $h(\epsilon_{ij})$ is normalized, i.e.  $\int_{-\infty}^\infty
d\varepsilon h(\varepsilon)=1,$ and it satisfies
$h(\varepsilon_{ij})=h(-\varepsilon_{ij})$.
For irrorganic amorphous semiconductors, $g(\varepsilon)$ is
approximated by the exponential function
\cite{SolidStateCommun.37.49,TMMA,MTM,PhysRevB.83.045201}, and its tail
is referred to as the Urbach tail \cite{PhysRev.92.1324}.
In this case, $h$ is the Laplace function.
For organic ones, $g(\varepsilon)$ is approximated by the Gaussian
function
\cite{JPhysChem.87.552,Bassler,PhysRevLett.91.216601,PhysRevB.47.4289},
and $h$ is also Gaussian.

\begin{figure}[htb]
\includegraphics[width=8.5cm]{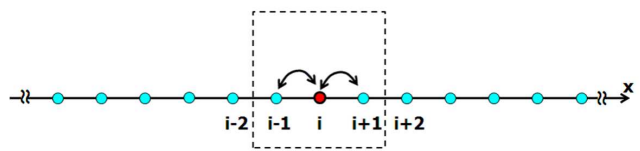} 
\caption{(Color online) A schematic of the disorder model.  It is
one-dimensional and a carrier can move only to either of the
first-nearest neighbors in a single hop.  }
\label{system}
\end{figure}

As for the hopping rate of the carriers, a realistic three-dimensional
model is somewhat complicated for theoretical considerations, and might
shadow the essence.
To elucidate the discussion, we adopt a simplified one-dimensional
model, where carriers can move only to either of the first-nearest
neighboring sites in a single hop (Fig.~\ref{system}).
This implies that we are focusing on a time scale where hopping to the
second-nearest neighboring sites is negligible.
We also assume that the number density of the carriers is small enough
so that the occupation of the states can be neglected.
With these assumptions, the hopping rate of the carrier from site $i$ to
site $i\pm1$, which we denote $\nu_{\pm}$, is approximately given by
\begin{align}
\nu_{\pm}
=
\nu_0 e^{-2a/\xi -(\varepsilon_{\pm}\pm Fa)\Theta(\varepsilon_{\pm}\pm Fa)/k_BT},
\label{MA}
\end{align}
where $a$ is the lattice spacing, $\xi$ is the localization length of
the localized state, $\nu_0$ is the typical magnitude of the hopping
rate, $\varepsilon_{\pm}\equiv \varepsilon_{i\pm1}-\varepsilon_i$,
$\Theta(x)$ is the Heaviside's step function, where $\Theta(x)=1$ for
$x>0$ and 0 otherwise, and $-F\ (F>0)$ is a constant external force
exerted on the carrier \cite{MA, AHL}.
%
%
%
%
For instance, $F$ can be a force due to an electric field
\cite{PhysRevB.86.045207}.
Note that the detailed balance is assumed in deriving Eq.~(\ref{MA}).

To describe the coarse-grained collective motion of the carriers, 
we consider a continuum model of the hopping conductance.
We assume that the hopping process is described by CTRW with the waiting
time density $\langle w (t)\rangle\sim\left(\alpha
A_\alpha/\Gamma(1-\alpha)\right) t^{-(1+\alpha)}$ $(0<\alpha<1)$ for
$t\to\infty$.
Here, $\Gamma$ is the Gamma function and $A_\alpha$ is a constant. 
This assumption is at least suitable for the exponential DOS
 \cite{ESW,ESW2,HBJK} and the Gaussian DOS \cite{HBJK}.
In fact, simple approximate analytic expressions for $\alpha$ can be
derived in terms of the microscopic parameters such as the width of the
DOS, both for the diffusive system \cite{ESW} and the system under a
constant external field \cite{ESW2}.
%
%
Then, by utilizing the mathematical technique presented in
Ref.~\cite{PhysRevE.61.132}, we can show that the probability density of
the carrier, $P(x,t)$, satisfies the following fractional
diffusion-advection equation (FDAE) in the continuum limit, i.e. $a\to
0$ and $A_{\alpha}\to 0$ with the generalized diffusion coefficient
$D_{\alpha} \equiv a^2/(2A_{\alpha})$ kept finite:
\begin{align}
\frac{\del P(x,t)}{\del t}
=
\ {}_0 \mathcal{D}_t^{1-\alpha}
\left(
\frac{\del }{\del x}
\mu_\alpha
+
D_\alpha\frac{\del^2}{\del x^2}
\right)P(x,t).
\label{FDAE}
\end{align}
Here, the operator $\ _0 \mathcal{D}_t^{1-\alpha}$ is the fractional
derivative, which is defined by the Riemann-Liouville operator, $\ _0
\mathcal{D}_t^{1-\alpha}P(x,t) =
\frac{1}{\Gamma(\alpha)}\frac{\del}{\del t}\int_0^t\dd
t'\frac{P(x,t')}{(t-t')^{1-\alpha}} $ \cite{OS},
and
\begin{eqnarray}
\mu_{\alpha}
=
\left[ 
\left\langle W_{+} \right\rangle
-
\left\langle W_{-} \right\rangle
\right] 
\frac{D_{\alpha}}{a}
\label{mobility}
\end{eqnarray}
is the generalized mobility, where $\langle W_{\pm}\rangle$ is the
coarse-grained hopping probability.
Note that $\langle W_{+}\rangle + \langle W_{-}\rangle=1$ holds by
definition.
It is natural to give $\langle W_{\pm}\rangle$ in terms of the bare
(microscopic) hopping probability $\nu_{\pm}$ by
\begin{eqnarray}
\langle W_\pm \rangle
\equiv
\langle \nu_\pm/(\nu_++\nu_-) \rangle,
\label{W_cg}
\end{eqnarray}
where $\langle \cdots \rangle = \int_{0}^{\infty} d\epsilon_{+}
h(\epsilon_{+}) \int_{0}^{\infty} d\epsilon_{-} h(\epsilon_{-}) \cdots$
is the average with respect to the distribution of the energy
difference.
It is straightforward to derive the following expression for $\langle
W_+\rangle$ from Eqs.~(\ref{MA}) and (\ref{W_cg}),
\begin{align}
\langle W_+\rangle
=&
\frac{1}{2}
-
\frac{1}{2}
\frac{\tilde{F}}{\tilde{T}}
\left[		
\frac{1}{2}
\int_{0}^{\infty} d\tilde{\varepsilon}_+		
\tilde{h}(\tilde{\varepsilon}_+)
\frac{1}{
	\cosh^2\left(\frac{\tilde{\varepsilon}_{+}}{2\tilde{T}}\right)
}
\right.
\nn
\\
&
\left.
+
\int_{0}^{\infty} d\tilde{\varepsilon}_+
\int_{0}^\infty d\tilde{\varepsilon}_-
\tilde{h}(\tilde{\varepsilon}_+)\tilde{h}(\tilde{\varepsilon}_-)
	\frac{1}{
	\cosh^2\left(\frac{\tilde{\varepsilon}_{+}-\tilde{\varepsilon}_{-}}{2\tilde{T}}\right)
}
\right]
\nn
\\
&+
\mathcal{O}\left( (\tilde{F}/\tilde{T})^2\right),
\label{EP}
\end{align}
%
%
\begin{figure*}[htb]
\includegraphics[width=8.25cm]{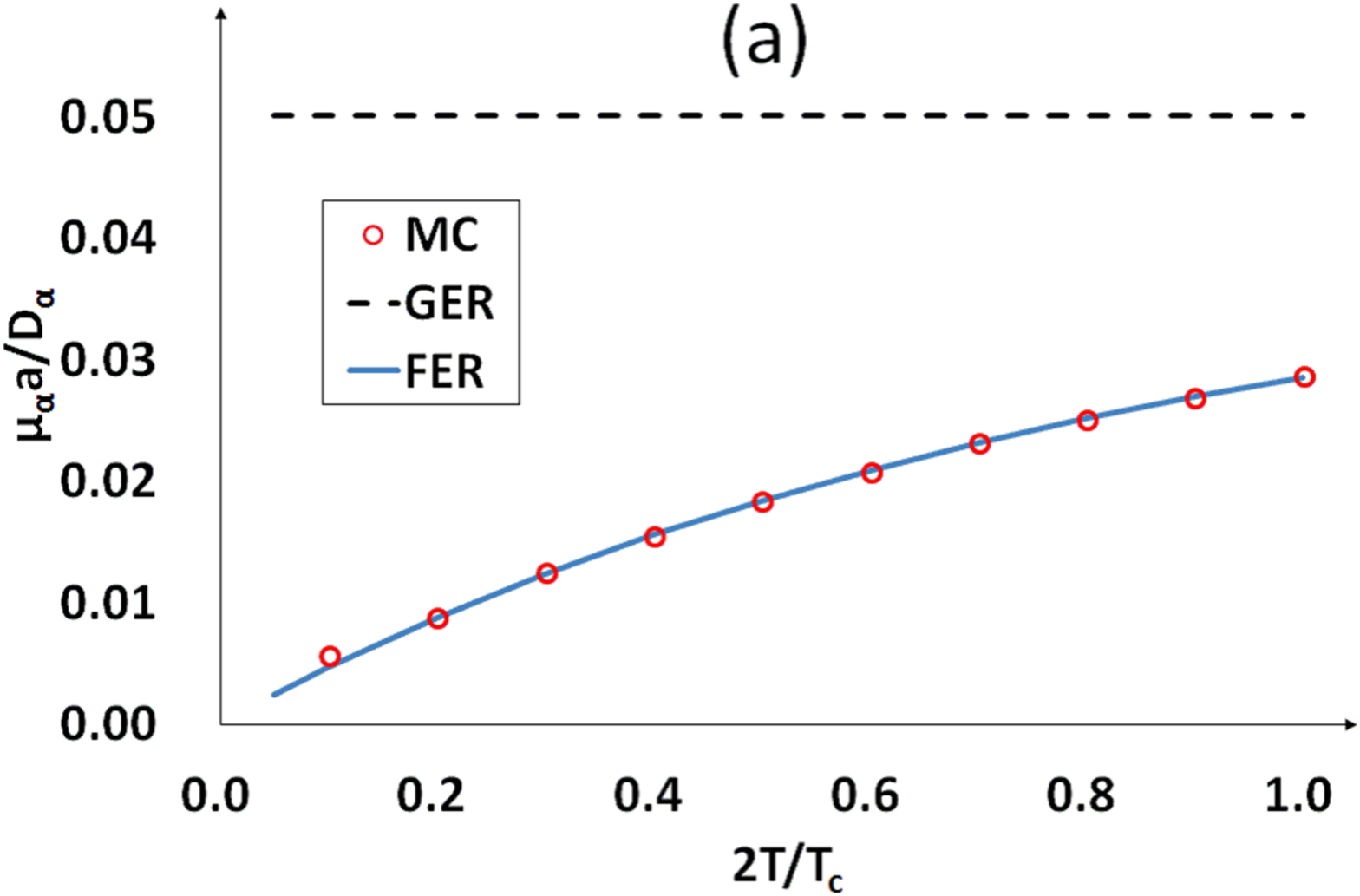}	
\hspace{3em}
\includegraphics[width=8.25cm]{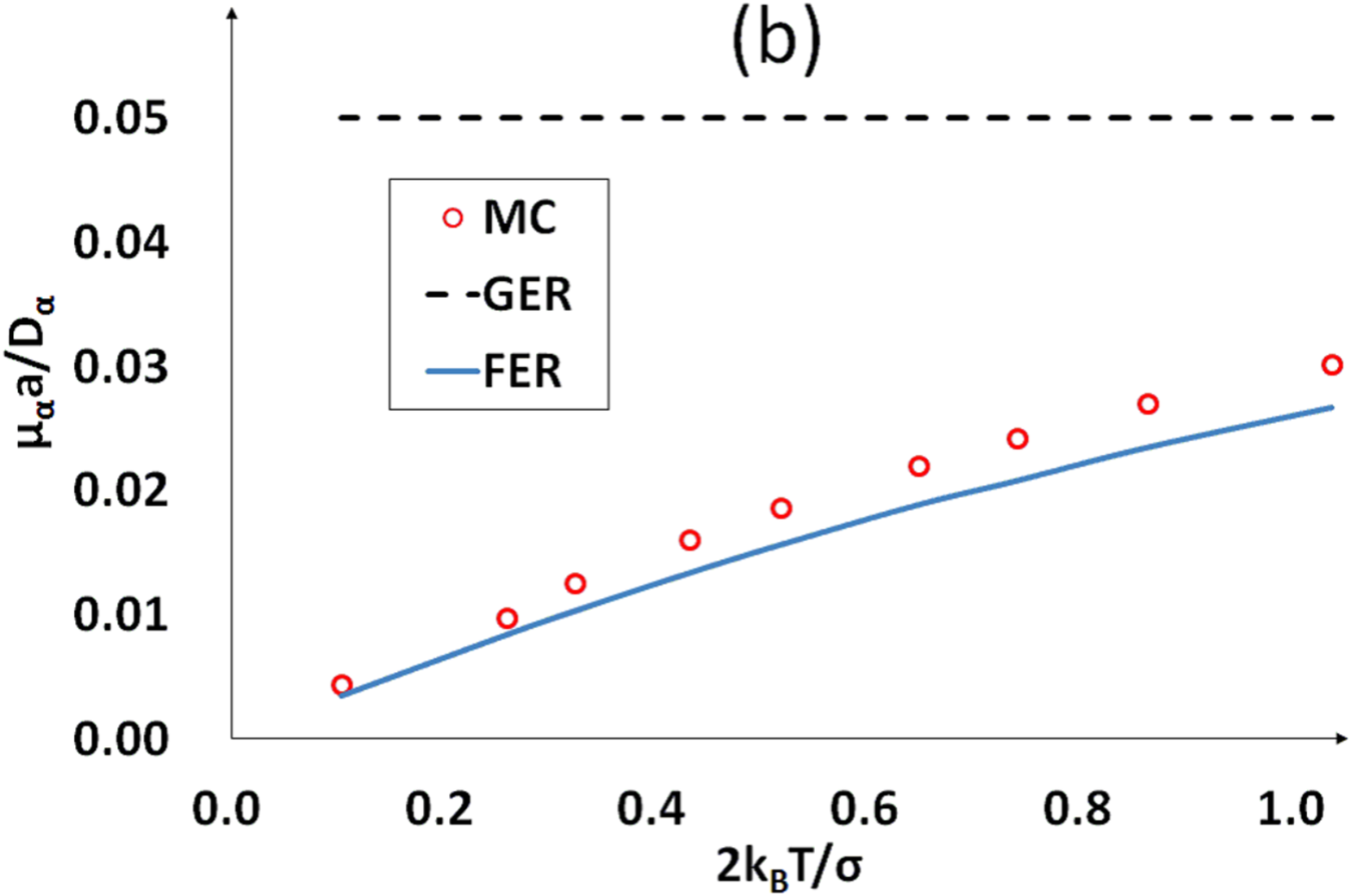}	
\caption{(Color online) Comparison of the fractional Einstein relation
(FER) and the generalized Einstein relation (GER) with MC
simulation, for the case of (a) exponential DOS and (b) Gaussian DOS.
The simulation conditions common for (a) and (b) are $N_P=10^6$, $a/\xi
= 10 $, $Fa/k_BT=0.1$, and (a) $2T /T_c = (0.1-1.0)$, (b)
$2k_BT/\sigma=(0.1-1.0)$.}
\label{Sim}
\end{figure*}
%
%
where the dimensionless variables are defined by
$\tilde{\varepsilon}_\pm=\varepsilon_\pm/\varepsilon_c$,
$\tilde{F}=Fa/\varepsilon_c$, $\tilde{T}=k_BT/\varepsilon_c$, and
$\tilde{h}\equiv \varepsilon_c h$, with $\varepsilon_c$ the typical
energy scale of the DOS.
Then, the following ER, which we refer to as the ``fractional Einstein
relation (FER)'', is obtained from Eqs.~(\ref{mobility}) and (\ref{EP}),
\begin{align}
\frac{\mu_\alpha a}{D_\alpha}
=&
2
\frac{\tilde{F}}{\tilde{T}}
\left[		
\frac{1}{2}
\int_{0}^{\infty} d\tilde{\varepsilon}_+		
\tilde{h}(\tilde{\varepsilon}_+)
\frac{1}{
\cosh^2\left(\frac{\tilde{\varepsilon}_{+}}{2\tilde{T}}\right)
}
\right.
\nn \\
&
\left.
+
\int_{0}^{\infty} d\tilde{\varepsilon}_+
\int_{0}^\infty d\tilde{\varepsilon}_-
\tilde{h}(\tilde{\varepsilon}_+)\tilde{h}(\tilde{\varepsilon}_-)
\frac{1}{
\cosh^2\left(\frac{\tilde{\varepsilon}_{+}-\tilde{\varepsilon}_{-}}{2\tilde{T}}\right)
}
\right]
\nn \\
&+
\mathcal{O}\left( (\tilde{F}/\tilde{T})^2\right).
\label{FER}
\end{align}
%

\vspace{1em}
{\it Simulation.--} 
The validity of Eq.~(\ref{FER}) is examined by MC simulation of the
hopping conductance, where the hopping probability is given by
Eq.~(\ref{MA}).
We consider the exponential DOS,
$g_{\mathrm{exp}}(\varepsilon)=e^{\varepsilon/k_BT_c}/k_BT_c$
($-\infty\leq \varepsilon\leq 0$), and the Gaussian DOS,
$g_{\mathrm{Gauss}}(\varepsilon)=e^{-\varepsilon^2/2\sigma^2}/\sqrt{2\pi\sigma^2}
$ ($-\infty\leq \varepsilon \leq \infty$), which are well established
for irrorganic and organic amorphous semiconductors, respectively.
Here, $k_BT_c$ and $\sigma$ are the typical widths of the DOS, which
correspond to $\varepsilon_c$ in the previous section.
The dimensionless DOS are given by
$\tilde{g}_{\mathrm{exp}}=e^{\tilde{\varepsilon}}$ and
$\tilde{g}_{\mathrm{Gauss}}= e^{-\tilde{\varepsilon}^2/2}/\sqrt{2\pi} $
, and the corresponding distribution functions of the energy differences
$\tilde{\varepsilon}_\pm$ ($-\infty\leq \tilde{\varepsilon}_\pm \leq
\infty$) are given by $\tilde{h}_{\mathrm{exp}}
=e^{-|\tilde{\varepsilon}_\pm|}/2$ and
$\tilde{h}_{\mathrm{Gauss}}=e^{-\tilde{\varepsilon}_\pm^2/4}/\sqrt{4\pi}
$, respectively.

The simulation method is the same as that in Refs.~\cite{ESW, ESW2}. 
The conditions are as follows:
the number of carriers (which is essentially the number of the trials of
the simulation performed) is $N_P=10^6$, and the parameters are chosen
as $a/\xi = 10 $, $2T /T_c = (0.1-1.0)$, $2k_BT/\sigma=(0.1-1.0)$, and
$Fa/k_BT=0.1$.
Note that the external force $F$ is constatly scaled with repsect to the
termperature $T$.
Both of the ranges, $2T /T_c = (0.1-1.0)$ and $2k_BT/\sigma=(0.1-1.0)$,
correspond to $0.1\leq \alpha \leq 1$.
Initially, all the carriers are rested at the origin. The generalized
mobility can be estimated from the relation $\langle x(t)\ran_F=
\mu_\alpha{t}^{\alpha}/\Gamma(\alpha+1)$.
The generalized diffusion coefficient can be estimated from the
mean-squared displacement of the carriers, $\lan x^2(t)\ran_F=
2\mu_\alpha^2 t^{2\alpha}/\Gamma(2\alpha+1) +
2D_\alpha{t}^{\alpha}/\Gamma(\alpha+1)$.
However, because the effective waiting time in the weak field is almost
the same as that in the diffusive case \cite{ESW2,PhysRevE.58.1296}, we
can estimate $D_\alpha$ from $\lan x^2(t)\ran_0=
2D_\alpha{t}^{\alpha}/\Gamma(\alpha+1)$. 
This also makes it possible to compare FER with GER directly.

In Fig.~\ref{Sim}, we show the comparison of FER and GER with MC
simulation. 
%
%
The result of MC simulation is sampled at the dimensionless time
$\tilde{t} \equiv \nu_0t=10^{14}$.  
The result of FER is obtained by performing the integrals in
Eq.~(\ref{FER}) numerically.
Fig.~\ref{Sim} (a) is the result for the exponential DOS, while
Fig.~\ref{Sim} (b) is that for the Gaussian DOS.
The horizontal axes are the dimensionless temperature, while the
vertical axes are the dimensionless ratio of the generalized mobility to
the generalized diffusion coefficient, $\mu_\alpha a/D_\alpha$.
The solid line, the dashed line, and the circles correspond to FER, GER,
and MC simulation.
From Fig.~\ref{Sim} (a) and (b), one can see that FER exhibits the
monotonically increasing tendency of $\mu_{\alpha} a/D_{\alpha}$ against
the dimensionless temperature, observed in MC simulation, which is
clearly beyond the reach of GER.
In addition, the quantitative agreement of FER and MC simulation is
surprisingly good for the case of exponential DOS, whereas the agreement
is less accurate for the the case of Gaussian DOS.

\vspace{1em}
{\it Discussion.--}\ 
In this section, we discuss the relation of our results to the previous
studies.
First of all, let us consider the limit $\epsilon_{c}\to 0$, which
corresponds to the high-temperature limit, $\tilde{T} =
k_{B}T/\epsilon_{c} \to \infty$.
In this case, the carriers are thermally excited up to the conduction
band, and hence the Ohmic conduction dominates the hopping conduction.
Theoretically, $\epsilon_{c}\to 0$ results in $\alpha\to 1$ and
$\tilde{h}(\tilde{\varepsilon}_\pm)\to\left[\lim_{\tilde{\varepsilon}_\pm\to+0}\delta(\tilde{\varepsilon}_\pm)+\lim_{\tilde{\varepsilon}_\pm\to-0}\delta(\tilde{\varepsilon}_\pm)\right]/2$,
which reduces Eq.~(\ref{FER}) to the original ER, $\mu a/D=
\tilde{F}/\tilde{T}$.
Hence, FER is an extension of the ER to the low-temperature regime,
where the hopping conduction cannot be neglected.

Next, we consider the case of finite $\epsilon_{c}$.
If we impose the detailed balance to the coarse-grained hopping process,
\begin{eqnarray}
\left\langle W_{+} \right\rangle
-
\left\langle W_{-} \right\rangle
=
\frac{Fa}{2k_{B}T} 
\end{eqnarray}
holds \cite{PhysRevE.61.132}.
Then, we obtain from Eq.~(\ref{mobility})
\begin{eqnarray}
\frac{\mu_{\alpha}a}{D_{\alpha}} 
=
\frac{\tilde{F}}{\tilde{T}},
\label{GER2}
\end{eqnarray}
which corresponds to GER.
In fact, the detailed balance is imposed to the coarse-grained model
such as CTRW in
GER~\cite{PhysRevE.58.1296,PhysRevE.63.046118,PhysRep.195.127}.
On the other hand, we have imposed the detailed balance to the
microscopic hopping process, where the bare hopping rate is given by
Eq.~(\ref{MA}).
In this case, it is notable that the detailed balance is violated by the
coarse graining, which is manifested in the coarse-grained hopping rate
$\lan W_\pm\ran$ given by Eq.~(\ref{EP}).
It is clear from Eqs.~(\ref{FER}) and (\ref{GER2}) that FER includes
corrections which depend on the microscopic details of the system, such
as DOS, while such a correction is absent in GER.
These corrections are crucial in describing the non-trivial dependence
of $\mu_{\alpha}a/D_{\alpha}$ on the dimensionless temperature, for the
case of constant $Fa/k_{B}T$.
Hence, we can see that the significant features of FER originate in
the violation of the detailed balance in the coarse-grained hopping
process.
Imposing the detailed balance to the coarse-grained hopping process,
which has been conventionally performed, neglects the essential features
of this process.

\begin{figure}[htb]
\includegraphics[width=8.0cm]{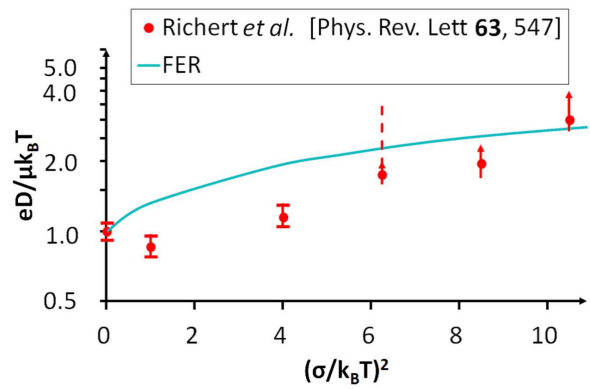}	
\caption{(Color online) Comparison of FER with the result of MC
 simulation of Ref.~\cite{PhysRevLett.63.547}.  
The parameter is $E=10^5$[V/cm]. 
}
\label{com}
\end{figure}

Finally, we compare FER with the result of MC simulation presented in
Ref.~\cite{PhysRevLett.63.547}.
In Ref.~\cite{PhysRevLett.63.547}, the hopping sites are distributed on
a three-dimensional cubic lattice with periodic boundary conditions, and
the external force $F$ is given by $F=eE$, where $e$ is the elementary
charge and $E$ is a constant electric field.
DOS is given by a Gaussian function with width $\sigma$.
The result of MC simulation of Ref.~\cite{PhysRevLett.63.547} is shown
by (red) circles with error bars in Fig.~\ref{com}, where the horizontal
axis is $(\sigma/k_BT)^2$ and the vertical axis is $eD/\mu k_BT$.
The upward arrows in Fig. \ref{com} show that the data become larger as
time goes on, which implies that the calculation performed is not long
enough for these conditions (probably due to limited calculational
resource at that time).
For $\sigma/k_BT=2.5$, we show the range of this time evolution with a
dashed line, which is obtained from the inset of Fig.~2 in
Ref.~\cite{PhysRevLett.63.547}.
The result of FER with the corresponding conditions is also shown by a
(blue) solid line in Fig.~\ref{com}.
We have estimated $eD/\mu k_BT$ by $D_\alpha\tilde{F}/\mu_\alpha a
\tilde{T}$, since the diffusion is dominant for the system under weak
external fields.
%
%
%
%
From Fig.~\ref{com}, we can see that both the result of
Ref.~\cite{PhysRevLett.63.547} and FER exhibit a monotonically
increasing tendency of $eD/\mu k_BT$ with respect to the inverse
dimensionless temperature.
Moreover, the quantitative agreement of the two results seem to be
relatively well, considering that FER is derived for a simplified
one-dimensional model.
Deriving FER for two- or three-dimensional models is a future work,
although we expect that the essential features of the hopping
conductance of non-crystalline semiconductors are already captured in
this letter by the simplified one-dimensional model.



%
\vspace{1em}
{\it Summary.--}\
In this study, we have presented a novel Einstein relation, which we
refer to as the ``fractional Einstein relation (FER)``, for the electric
conduction in non-crystalline semiconductors.
FER is derived from the fractional diffusion-advection equation (FDAE),
together with coarse-graining the bare (microscopic) hopping probability
where the detailed balance is imposed.
The striking feature of FER is that it includes microscopic properties
such as the probability distribution of the energy difference of the
hopping sites, which can be obtained from DOS.
This is not the case for the original ER, nor for the generalized
Einstein relation (GER).
It has been shown by comparing with MC simulation that the dependence of
FER on DOS is essential to reproduce the non-trivial dependence of the
ratio of the mobility to the diffusion coefficient on the dimensionless
temperature.
The crucial difference between FER and the original ER, or GER, is that
the detailed balance of the coarse-grained hopping process is violated
in FER, while it is preserved in others.
This indicates that the violation of the detailed balance of the
coarse-grained hopping process is a key feature for the electric
conduction in non-crystalline semiconductors.

\vspace{1em}
{\it Acknowledgements}.
We are grateful to K. Shinjo, K. Nagane, and the members of the Analysis Technology Development Department 1 for 
fruitful discussions and support.

\bibliography{99_refs.bib}	

\end{document}